# Substrate matters: Surface-polariton enhanced infrared nanospectroscopy of molecular vibrations

Marta Autore[1], Lars Mester[1], Monika Goikoetxea[1], R. Hillenbrand[1,2]
[1] CIC nanoGUNE, 20018 Donostia-San Sebastián, Spain
[2] IKERBASQUE, Basque Foundation for Science, 48013 Bilbao, Spain
r.hillenbrand@nanogune.eu

**Infrared nanospectroscopy based on Fourier transform infrared near-field spectroscopy (nano-FTIR) is an emerging nanoanalytical tool with large application potential for label-free mapping and identification of organic and inorganic materials with nanoscale spatial resolution. However, the detection of thin molecular layers and nanostructures on standard substrates is still challenged by weak signals. Here, we demonstrate a significant enhancement of nano-FTIR signals of a thin organic layer by exploiting polariton-resonant tip-substrate coupling and surface polariton illumination of the probing tip. When the molecular vibration matches the tip-substrate resonance, we achieve up to nearly one order of magnitude signal enhancement on a phonon-polaritonic quartz (c-SiO$_2$) substrate, as compared to nano-FTIR spectra obtained on metal (Au) substrates, and up to two orders of magnitude when compared to the standard infrared spectroscopy substrate CaF$_2$. Our results will be of critical importance for boosting nano-FTIR spectroscopy towards the routine detection of monolayers and single molecules.**

Scattering-type scanning near-field optical microscopy (s-SNOM)[1] and Fourier transform infrared nanospectroscopy (nano-FTIR)[2,3,4] are rapidly emerging techniques that allow for infrared characterization of materials with nanoscale spatial resolution. During the past years, they have been widely applied in numerous studies of both fundamental and applied aspects, including nano-FTIR chemical nanoidentification of organic and inorganic samples[3,2,4,5], contact-less s-SNOM conductivity and phase transition mapping[2,6,7,8,9,10] and secondary structure mapping of protein complexes[11]. However, when applied to very thin layers of weakly absorbing substances, particularly monolayers or individual molecular complexes, s-SNOM and particularly nano-FTIR signals are often at or below the detection limit[11,12,13].

s-SNOM and nano-FTIR spectroscopy are scanning near-field optical probe techniques based on illuminating an atomic force microscope (AFM) tip with typically p-polarized laser radiation. The tip acts as an antenna and concentrates the incident field into a nanoscale near-field hotspot at its apex. Local near-field interaction between tip and sample modifies the tip-scattered light. Recording of the tip-scattered light as function of tip position subsequently yields nanoscale-resolved images of local optical properties of the sample. In nano-FTIR, broadband infrared illumination is utilized and the tip-scattered light is analysed with an asymmetric Fourier-transform spectrometer to obtain infrared amplitude and phase spectra that in case of low-index materials correspond to local reflectivity and absorption spectra, respectively. The spatial resolution is determined essentially the tip radius, which typically is in the order of a few ten nanometers.

To enhance the infrared near-field signals in s-SNOM and nano-FTIR, the sample layers are typically deposited on strongly reflecting substrates such as silicon or gold[1,3,4,5,12,13,14,15,16,17]. The signal enhancement can be explained by the electromagnetic field concentration in the gap between tip and substrate, similar to tip-enhanced Raman spectroscopy (TERS) in gap mode configuration[18]. However, despite various recent developments and rapid diffusion of the nano-FTIR technique, only few studies discuss in detail the signal enhancement by tip-substrate coupling and its potential further exploitation beyond standard Si and Au substrates. Recently, spectroscopic s-SNOM imaging of molecules deposited on infrared-resonant metal antennas has demonstrated an enhancement of the vibrational signature[19,20-21]. However, this approach requires the fabrication of antennas and limits the enhancement to the relatively small areas where electromagnetic hot spots are formed. The predicted enhancement of s-SNOM and nano-FTIR spectroscopic signatures of weak molecular vibrations via tip-induced polariton excitation in flat substrates[17] has not been demonstrated so far experimentally.

Here we study the influence of the substrate on nano-FTIR molecular vibrational spectroscopic features and provide first experimental evidence that they can be enhanced by one to two orders of magnitude when placing them on flat polaritonic substrates instead of standard IR substrates such as $CaF_2$, provided that the molecular vibration matches the tip-substrate resonance. Particularly, we study 13 nm thin layers of poly(ethylene oxide) (PEO) molecules and demonstrate experimentally that a significant signal enhancement - beyond that provided by flat gold substrates - can be achieved due to polariton-resonant near-field coupling between tip and substrate, here a quartz crystal featuring a tip-induced phonon polariton resonance matching the molecular vibration of PEO. An even higher signal enhancement can be achieved by additional illumination of the tip via surface phonon polaritons on the quartz crystal that are launched at discontinuities on the sample surface.

Nano-FTIR spectroscopy was performed with the setup[3] illustrated in Figure 1a (neaSNOM from Neaspec GmbH). A broadband mid-infrared supercontinuum (SCIR) laser pulse (650-1450 cm$^{-1}$) generated by a difference frequency generation laser source (FemtoFiber dichro midIR, Toptica) is focused with a parabolic mirror (NA = 0.4) onto a commercial metallized AFM tip (Arrow NCPt, Nanoworld), after passing through a beam splitter (BS). The light backscattered by the tip is collected with the same parabolic mirror and directed together with the reference beam to an MCT mid-infrared detector (IRA-20-00103, InfraRed Associates). At a fixed tip and sample position, the reference mirror is translated to record an interferogram. Fourier transform of the latter yields both amplitude and phase spectra of the tip-scattered field (note: both amplitude and phase are obtained, as the sample is located in one of the interferometer arms, in contrast to standard FTIR spectroscopy where the sample is placed outside the interferometer). To obtain background-free nano-FTIR spectra, the AFM tip is operated in tapping mode, i.e. it is oscillating vertically near the cantilever's mechanical resonance frequency $\Omega$, and the detector signal is demodulated at higher harmonics $n$ of $\Omega$. Each sample spectrum is normalised to a reference spectrum measured on a clean Au surface, yielding the final nano-FTIR amplitude and phase spectra $s_n/s_n(\text{Au})$ and $\varphi_n - \varphi_n(\text{Au})$, respectively. The studied samples consist of 13 nm-thick poly(ethylene oxide) layers on top of $CaF_2$, Si, Au and c-$SiO_2$ substrates. To that end, a 0.25% solution of PEO in acetonitrile was spincoated at 150 rps during 2 minutes on each substrate.

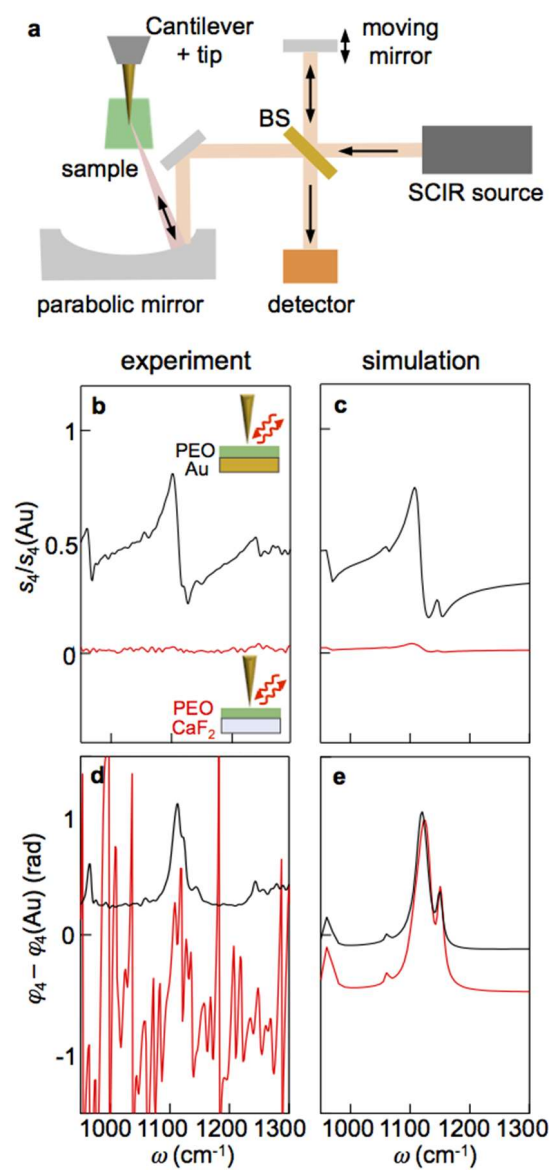

**Figure 1. nano-FTIR spectroscopy of 13 nm thick PEO layers on $CaF_2$ and Au substrates.** (a) Schematics of the nano-FTIR setup. (b,d) Experimental amplitude (b) and phase (d) spectra of 13 nm-thick PEO on Au (black curve) and $CaF_2$ substrate (red curve). (c,e) Calculated spectra, analogous to panels b,c. All spectra are normalized to the nano-FTIR spectrum of a bare Au sample.

Fig. 1b and d show the measured amplitude and phase spectra of PEO on Au and $CaF_2$, respectively. For PEO on $CaF_2$ (red curves) we observe a near-zero amplitude signal with a barely visible spectral feature at around 1105 cm$^{-1}$, which can be attributed to the symmetric and asymmetric stretching of the C-O-C group of PEO[22]. It is better seen in the phase spectrum, although the signal-to-noise-ratio (S/N) is rather low. Strikingly, for PEO on Au we obtain a typical dispersive-like feature in the amplitude spectrum (black curve in Fig. 1b), where the signal difference between maximum and minimum is a factor of about 20 larger than for PEO on CaF$_2$. Further, the phase spectrum (black curve in Fig. 1d) exhibits a clear peak with a significantly improved S/N ratio as compared to the PEO spectrum on CaF$_2$.

We corroborate the experimental nano-FTIR spectra and the substrate-induced signal enhancement with calculations using the multilayer finite dipole model, in which the tip is described as an elongated dipole[23] that interacts with the several sample layers[24] that are characterized by their dielectric function. In this model, the tip-scattered field is given by

$$E_s \propto (1+r)^2 \alpha_{\text{eff}} \qquad (1)$$

where $\alpha_{\text{eff}}$ is the effective polarizability of the tip that accounts for the near-field interaction between the tip and the layered sample. $r$ is the far-field reflection coefficient of the substrate (for calculation details see Methods). The calculated amplitude and phase spectra are shown in Figure 1c and 1e. We observe a remarkable agreement between experiment and calculations, particularly regarding the strong enhancement of the vibrational signature in the amplitude spectrum of the PEO layer on the Au substrate as compared to PEO on CaF$_2$. This signal amplitude enhancement is caused by (*i*) the large far-field reflection coefficient of Au ($r$ = 1) as compared to CaF$_2$ ($r$ = -0.06), which is the Fresnel reflection coefficient obtained for a p-polarized wave at 60 degree incidence normal to the sample surface and a refractive index $n_{\text{CaF2}}$ = 1.34)[25] which leads to a more efficient tip illumination and collection of the tip-scattered field, and (*ii*) the increased near-field interaction between tip and gold substrate yielding to a stronger local field enhancement at the tip apex[14,17]. On the other hand, the calculated phase spectra indicate that the phase contrast (peak height) of PEO on CaF$_2$ is slightly larger than that of PEO on Au. However, because of the much lower signal on CaF$_2$ (see amplitude spectra), the experimental S/N ratio is too low for this effect to be beneficial in a practical way.

In Figure 2 we explore the possibility of further increasing the nano-FTIR signals of PEO by resonant tip-substrate coupling. We recall that in most models for tip-sample near-field interaction the effective polarizability of the tip, $\alpha_{\text{eff}}$, is a function of the near-field (quasi-electrostatic) reflection coefficient $\beta$, which for a semi-infinite surface in vacuum is given by

$$\beta = \frac{(\varepsilon - 1)}{(\varepsilon + 1)} \qquad (2)$$

where $\varepsilon$ is the dielectric function of the sample[1,3,17,26]. The effective polarizability, and thus the tip-scattered field, exhibit a resonance near $\varepsilon \approx -1$ owing to the divergence of $\beta$, which corresponds to the excitation of localized surface polaritons by the concentrated near fields at the tip apex (e.g. surface plasmon polaritons in metals or doped semiconductors or surface phonon polaritons in polar crystals[26]. In Ref. 17 we predicted that this resonant near-field interaction could enhance the infrared spectral contrast of molecular vibrations of thin layers in the tip-substrate gap. Up to date, however, this effect has not been demonstrated experimentally.

For a first experimental demonstration of surface-polariton enhanced nano-FTIR spectroscopy of molecular vibrations, we selected crystalline quartz (c-SiO$_2$) substrates, an exemplary polar crystal supporting surface phonon polaritons[27,28] near the molecular vibrational resonance of PEO. The measured far-field reflectivity spectrum of c-SiO$_2$ is shown in Figure 2a. It exhibits a typical *Reststrahlen band*, where $R \approx 1$. Within this band (between 1070 and 1220 cm$^{-1}$), the real part of the dielectric function, $\varepsilon_1$, is negative, which in case of c-SiO$_2$ is a consequence of multiple phonon excitations. By fitting the far-field spectrum with Fresnel equations we obtained the complex-valued dielectric function of c-SiO$_2$, which is plotted in Fig. 2b. Around 1150 cm$^{-1}$ we find $\varepsilon_{1,c-SiO} \simeq -1$, such that the condition for a tip-induced polariton resonance in the substrate is fulfilled near the PEO molecular vibrational resonance at 1105 cm$^{-1}$ (marked by vertical dashed line and defined by the maximum of $\varepsilon_{2,PEO}$) that can be clearly recognized in the dielectric function of PEO (Fig. 2c). Indeed, the nano-FTIR spectrum of a clean quartz surface (normalized to the spectrum of bare gold) reveals the expected resonance[27] (blue curve in Fig. 2c) where the normalized amplitude signal $s_4/s_4$(Au) exhibits values that are up to 3.5 (i.e. 3.5 times higher than that of the gold reference surface). The multiple-peak structure of the spectrum (maxima at 1135 and 1170 cm$^{-1}$) can be ascribed to the presence of several phonon resonances contributing to the Reststrahlen band of quartz[27].

To explore the effect of tip-quartz coupling on the nano-FTIR spectra of thin PEO layers, we evaporated 30 nm of Au onto one half of the quartz substrate and spin-coated PEO as described above. That way we obtained a PEO layer on gold and on quartz within one sample, which allows for a reliable comparison of the two situations under the same experimental conditions (same tip, as well as same microscope settings and adjustment).

We obtained self-assembled islands of PEO on top of the gold and quartz surfaces with a rather homogeneous height of 13 nm and an average diameter of about 150 nm. The PEO areas are clearly visible in the AFM image (1x0.5 μm$^2$) shown in Figure 2d and in the s-SNOM image of the same area taken at 1114 cm$^{-1}$ (amplitude image, $s_4$, Figure 2e). The formation of islands offers the advantage that nano-FTIR spectra of adjacent regions of PEO on c-SiO$_2$ and bare c-SiO$_2$, as well as of PEO on Au and bare Au, can be measured under the same experimental conditions, thus allowing for a reliable comparison of spectra. Note that in both data analysis and simulations we do not consider the island-like morphology of the PEO layer. The measurements were done in the center of larger islands, which have a diameter of typically 100-200 nm, and thus can be considered in first approximation as an extended layer compared to the tip of about 25 nm apex radius and corresponding spatial resolution.

In Figure 2f we compare nano-FTIR spectra of PEO on quartz with nano-FTIR spectra of pure quartz. We observe a strong dip in the PEO spectrum at the position close to the peak maximum of the pure quartz spectrum, which we attribute to the vibrational resonance of PEO (marked by dashed line). This finding strongly resembles the features found in surface- and antenna-enhanced infrared spectra (SEIRA) of molecules on resonant plasmonic antennas, where Fano interference between the molecular vibration and plasmons occurs[29,30]. Analogously, we can explain the observation in our nano-FTIR spectra by Fano interference between the molecular vibration and the tip-induced phonon-polariton in the substrate, similar to phononic SEIRA[31].

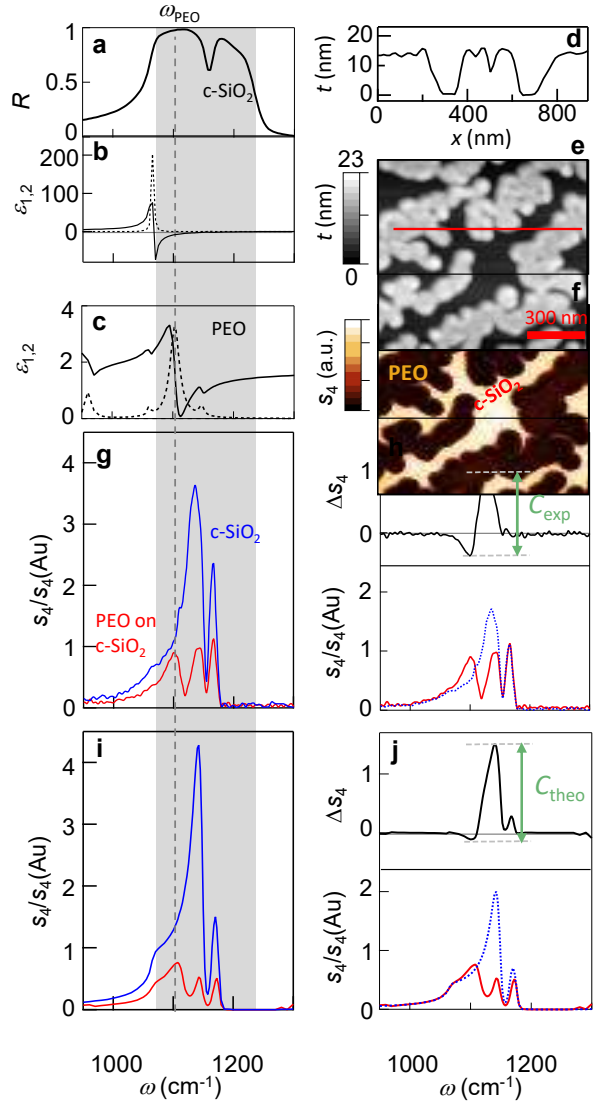

**Figure 2. nano-FTIR spectroscopy on a resonant quartz substrate.** (a) Reflectivity spectra of bare quartz. (b) Real (black curve) and imaginary (dashed black curve) part of the dielectric function obtained by fitting the Drude-Lorentz model to the quartz reflection spectrum using the RefFIT software[32]. (c) Real (black curve) and imaginary (dashed black curve) part of the dielectric function of PEO, obtained by fitting far-field transmission spectra of 500 nm thick PEO layers on $CaF_2$ by the Drude-Lorentz model using the RefFIT software. The gray shaded area indicates the Reststrahlen band of quartz and the vertical dashed line indicates the resonance frequency of the PEO vibration. (d) Topography line profile along the red line in panel e). (e) Topography and (f) s-SNOM image recorded at $\omega = 1114$ cm$^{-1}$ (shown for illustrative purposes only) of a 1x0.5 μm$^2$ c-SiO$_2$ substrate with spin-coated PEO on top, the latter forming nanoscale islands. (g) Experimental and (i) calculated nano-FTIR amplitude spectra (4$^{th}$ harmonics, normalized to gold spectrum) of quartz (blue) and PEO on quartz (red). Panels (h) and (j) show the normalized quartz spectra (blue dashed curves) used for calculating the maximum spectral contrast $C$ (defined as difference between maximum and minimum of $\Delta s_4$ shown by black curves) of the polariton-enhanced vibrational feature of PEO on quartz for experimental and calculated data, respectively.

We measure the magnitude of the polariton-enhanced molecular vibrational signature analogously to what is conventionally done in SEIRA measurements. We first subtract the spectrum of PEO on quartz from the spectrum of bare quartz and then calculate the difference between maximum and minimum of the resulting spectral feature (as illustrated in Figure 2g). To apply this procedure, we need to take into account the reduced tip-substrate coupling caused by the increased tip-substrate distance when the tip is on top of PEO. To that end, we normalized the spectrum of quartz, $s_4(\text{c-SiO}_2)/s_4(\text{Au})$, in such a way that the signal between 980-1050 cm$^{-1}$ (assumed to be barley affected by the PEO vibration because of its rather large spectral distance to the phonon resonance) matches with the one of the spectrum of PEO on quartz, $s_4(\text{PEO on c-SiO}_2)/s_4(\text{Au})$. We denote the resulting spectrum $\bar{s}_4(\text{c-SiO}_2)/s_4(\text{Au})$ and show it as blue dashed line in Figure 2d. With $\Delta s_4 = \bar{s}_4(\text{c-SiO}_2)/s_4(\text{Au}) - s_4(\text{PEO on c-SiO}_2)/s_4(\text{Au})$ we finally obtain the polariton-enhanced vibrational spectral contrast (black curve in Fig 2g). Although our normalization step does not take into account the spectral shifts caused by the increased tip-substrate distance[33,34] we can reliably determine the enhancement of $C$ on c-SiO$_2$ as compared to Au (see Supporting Information S1).

The spectral contrast $\Delta s_4$ exhibits an asymmetric shape, neither resembling the amplitude nor the phase spectrum of PEO on CaF$_2$ or Au substrates. This observation is typical for a Fano-type interaction between a broad excitation (the phonon polaritons in this case) and a narrow mode (the PEO vibration), where the spectral shape depends on both the interaction strength and on the spectral overlap between the two modes[29]. Most importantly, the spectral contrast $\Delta s_4$ is larger than the one obtained for PEO on Au, as we will highlight and quantify below. Before, we corroborate the experimental results by calculations using the multilayer finite dipole model described above. As shown in Figure 2h,i, the spectral behavior is well reproduced and the enhancement of the vibrational contrast is confirmed. We note that the near-field phase contrast $\Delta\varphi_4$ (shown in the Supporting Information S2) is barely enhanced. For that reason, we analyze in this work essentially near-field amplitude spectra.

In the following, we compare quantitatively the enhancement factors of the PEO vibrational signature in the case of non-resonant substrates with different reflection coefficients (CaF$_2$, Si, Au) and resonant c-SiO$_2$. The results are summarized in Figure 3b. The three top curves show the amplitude spectra of PEO on CaF$_2$ (green), Si (orange) and Au (black). Taking the maximum spectral contrast of the PEO on the Au substrate ($C$ = 0.46) as a reference, we find a reduction of $C$ to 54% for PEO on Si and to 7% for the CaF$_2$ substrate. This finding can be explained by (i) the reduced reflection coefficient $r$ of Si and CaF$_2$ compared to Au ($r_{\text{Si}}$ = 0.28, $r_{\text{CaF2}}$ = -0.06, $r_{\text{Au}}$ = 1)[25] which reduces both the illumination and scattering of the tip via the substrate (top illustration of Fig. 3a), and (ii) the reduced near-field interaction between tip and substrate owing the smaller near-field reflection coefficient $\beta$ (eq. 2) of Si and CaF$_2$[35]. When PEO is placed on c-SiO$_2$ we achieve a 300% increase (compared to Au substrate) of $C$, which is determined according to the procedure described above from the red spectrum shown as inset. We can explain this significant enhancement compared to Au only by the enhanced near-field interaction (described by $\alpha_{\text{eff}}$) between tip and substrate due to tip-induced phonon polariton excitation, as tip illumination and tip scattering via the substrate reaches is maximum for the Au surface (we stress that $r$ is the far-field Fresnel reflection coefficient of the substrate, which cannot exeed $r$ = 1). Actually, the tip illumination and tip scattering via the quartz surface reduces the signal amplitude by Abs[$(1+r_{\text{c-SiO2}})^2/(1+r_{\text{Au}})^2$]~50%, which is, however, more than compensated by the enhanced near-field interaction. Note that a native oxide is present on the Si substrate, with a phonon-resonance at the position of the PEO vibrational mode. However, this oxide layer is typically as thin as a few nanometer. Although it can be observed in s-SNOM, it provides a rather small and thus negligible contribution to the near-field interaction between tip and silicon substrate[36].

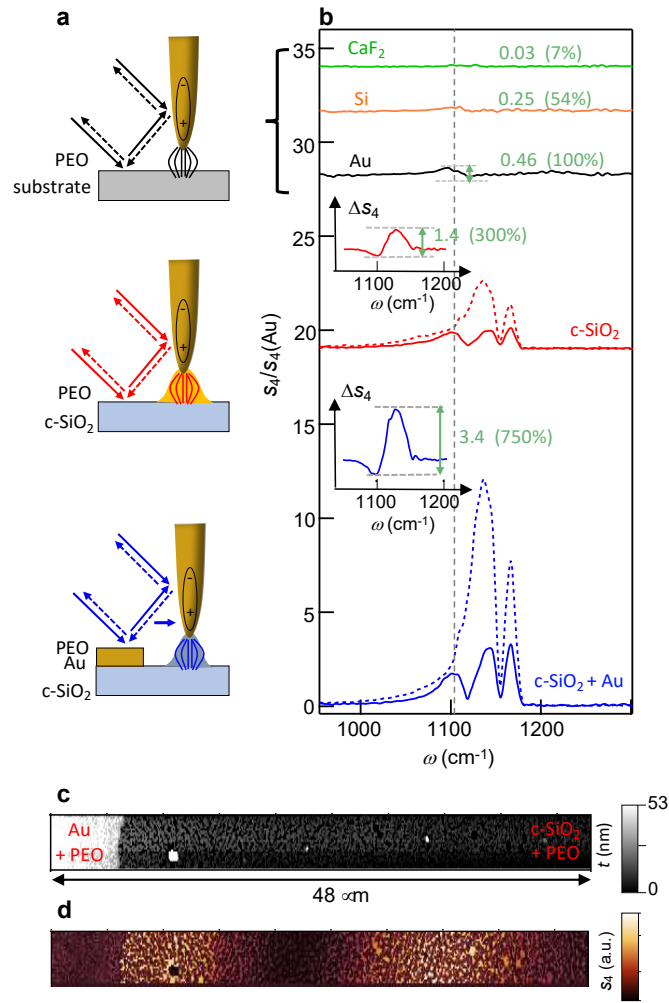

**Figure 3. Comparison of different substrates for nano-FTIR spectroscopy of PEO.** (a) Sketches of the different mechanisms contributing to the tip illumination in the performed experiments. From top to bottom: indirect illumination of the tip via the substrate (top), additional resonant tip-substrate coupling due to tip-induced phonon polariton excitation in the substrate (middle), and additional tip illumination via propagating surface phonon polaritons launched by the edge of a gold film (bottom). (b) From top to bottom: nano-FTIR amplitude spectra of PEO on $CaF_2$, Si, Au and quartz substrates (solid curves). The very bottom spectrum was obtained in 1 μm distance to the edge of an extended Au film launching surface phonon polaritons. Dotted lines show spectra without PEO. For better visibility of the vibrational features, the spectra on the different substrates are offset. Insets: The red and blue curves show the calculated spectral contrast $\Delta s_4$. Note that the scale of the vertical axes $\Delta s_4$ and $s_4/s_4(Au)$ are the same. (c) AFM topography and (d) s-SNOM infrared amplitude images of a large area of PEO on quartz close to the edge of a gold film. The infrared image was recorded at 1114 cm$^{-1}$ provided by a quantum cascade laser (MIRcat, Daylight Solutions).

We finally explore the possibility of further boosting the enhancement of the vibrational contrast by exploiting additional tip illumination via propagating surface polaritons. To that end, we deposited an extended gold film on quartz. The gold edge serves as a surface phonon polariton (SPhP) launcher[37] (see sketch in Fig. 3 panel a, bottom). When the tip is placed close to the gold edge, i.e. within the propagation length of the SPhPs, the polariton field provides an additional (coherent) tip illumination, thus boosting the near-field enhancement at the tip apex. The experimental verification of this concept is shown by the bottom spectra (blue) of Fig. 3b. The dashed blue curve represents the spectrum of quartz taken at 1 μm distance from the gold edge. The signal amplitude is up to 12 times higher than the signal on gold, and approximately 3 times higher than the signal of quartz far from the Au edge. We explain this significant additional signal enhancement by (*i*) the high reflection coefficient r of Au (which – as explained above – increases the signal amplitude by a factor of 2 compared to quartz) and (*ii*) the additional tip illumination through the SPhP (both aspects illustrated in the bottom sketch of Fig 3a). We estimate that the additional tip illumination through the SPhP enhances the signal amplitude by an additional factor of about 1.5.

To verify that the gold edge indeed launches a surface phonon polariton, we recorded a large s-SNOM image (Fig. 3c,d) at 1114 cm$^{-1}$ (photon wavelength $\lambda = 9.0$ μm) provided by a quantum cascade laser (MIRcat, Daylight Solutions). We see fringes parallel to the gold edge with a spacing of $d = 24$ μm, caused by the interference of the SPhP field with the incident field at the position of the tip apex[37,38]. For $\alpha = 42°$ (angle of incidence with respect to the sample surface) we obtain a SPhP wavelength of $\lambda_{\text{SPhP}} = 8.1$ μm according to $\lambda_{\text{SPhP}} = \lambda d/[\lambda + d \cos \alpha]$, matching the calculated SPhP wavelength using the relation $2\pi/\lambda_{\text{SPhP}} = \text{Re}[\omega/c \sqrt{\varepsilon_{c-S} / (\varepsilon_{c-\text{SiO}} + 1)}]$. Note that for this s-SNOM experiment the laser focus size had to be enlarged significantly as compared to the nano-FTIR experiments, in order to illuminate both tip and gold edge for distances as large as 100 μm, which is required for unambiguous observation of the interference fringes.

We apply the edge-launched SPhPs for enhancing nano-FTIR signals and measure the spectrum of PEO on quartz at 1 μm distance from the gold edge (solid blue curve in Fig. 3b). By determining the spectral contrast $\Delta s_4$ (blue curve, inset of Fig. 3b) we obtain $C = 3.4$, which corresponds to 750% of the value obtained for PEO on the Au substrate and almost 1400% of the value obtained on Si, the latter being the standard nano-FTIR substrate. As shown in the Supporting Information S3, the value of $C$ oscillates with distance to the gold edge according to the local field at the tip apex that is governed by the interference between the incident and SPhP field (Fig. 3d).

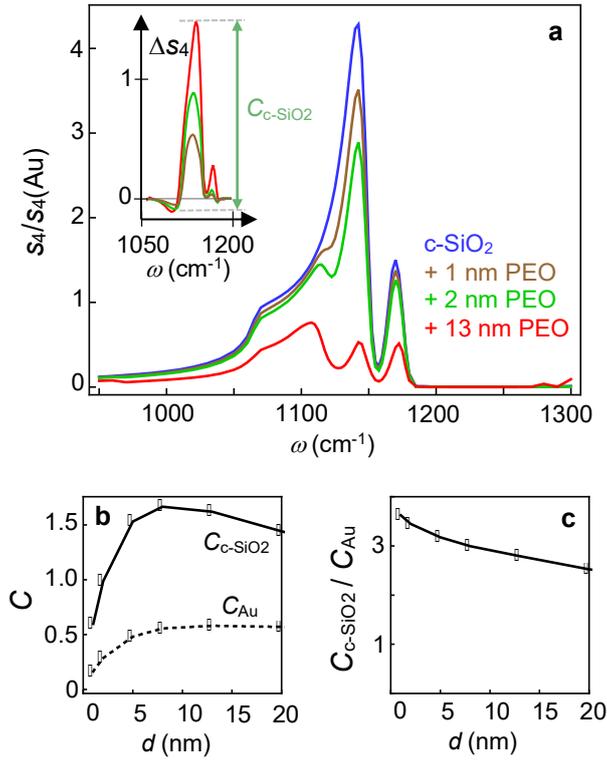

**Figure 4. Simulation of thickness-dependent spectra and spectral contrasts.** (a) Calculated nano-FTIR amplitude spectra of differently thick PEO layers on a quartz substrate. Inset: Spectral contrast $\Delta s_4$ of the differently thick PEO layers, analogous to Fig 2. (b) Maximum spectral contrast for differently thick PEO layers on quartz and Au substrates, $C_{\text{c-SiO}_2}$ and $C_{\text{Au}}$. (c) Ratio $C_{\text{c-SiO}_2}/C_{\text{Au}}$.

To gain further insights into the spectral contrast enhancement, particularly for thinner molecular layers, we show simulations of nano-FTIR spectra in Fig. 4 (experimentally we could not achieve PEO layers thinner than 13 nm). Fig. 4a shows spectra of 1 nm and 2 nm thick PEO layers on the quartz substrate (brown and green curves, respectively) in comparison to the 13 nm thick PEO layer on quartz (red curve) and the pure quartz spectrum (blue curve). We clearly see a significant modification of the quartz spectrum caused by even these thin PEO layers. Following the procedure described in Fig. 2, we calculate the maximum spectral contrast $C_{\text{c-SiO}_2}(d)$ of the molecular vibration of a PEO layer of thickness $d$ on a quartz substrate. The black curve in Fig. 4b shows that for a few nanometer thick layers the spectral contrast $C_{\text{c-SiO}_2}$ is larger than that of a 13 nm thick PEO layer on a gold substrate. $C_{\text{c-SiO}_2}$ reaches a maximum around 10 nm PEO thickness, before it decreases with increasing $d$. We explain this decrease by the decreasing tip-substrate near-field coupling. For a comparison, we show the thickness-dependent spectral contrast $C_{\text{Au}}(d)$ for PEO layers on a gold substrate (black dashed curve), highlighting that the enhancement of the PEO spectral contrast on the quartz substrate increases for thinner PEO layers. This observation can be better appreciated by plotting the ratio $C_{\text{c-SiO}_2}(d)/C_{\text{Au}}(d)$ in Fig. 4c (black curve). The increasing enhancement for thin layers can be explained by the stronger surface phonon polariton excitation when the distance between tip and quartz decreases.

The enhancement of nano-FTIR spectroscopy by tip-induced resonant surface polariton excitation could be realized with various substrates supporting either surface phonon polaritons (polar crystals such as SiC or $Al_2O_3$)[39], dielectrically loaded polar crystals that provide stronger surface polariton confinement[28,40] or surface plasmon polaritons. The latter exhibit resonances at mid-infrared frequencies on highly doped semiconductors, offering the advantage that the resonances can be spectrally shifted via the carrier to match specific molecular vibrations[7]. Strong resonances could be achieved with semiconductor with large carrier mobilities, such as GaAs or InAs[41,42,43,44]. On the other hand, tip-induced resonant surface polariton excitation has not been achieved with extended flat slabs (flakes) of uniaxial 2D materials (such as h-BN[45]) as a substrate, as they do not support *surface* polaritons (such as the ones employed in this work) at the slab surface. Surface polaritons exist at the flake edges[46] but resonant coupling with the tip has not been reported yet. Nevertheless, it might be worth exploring mechanisms to enhance nano-FTIR signals by exploiting graphene plasmons[47,48] or hyberbolic plasmon and phonon polaritons in multilayer 2D materials such as h-BN[45].

We finally note that focusing of substrate polaritons onto the probing tip – independent of polariton-resonant tip-substrate coupling - could further boost the nano-FTIR signals. This could be accomplished, for example, with appropriately designed polariton launchers on conventional polar crystals[49], metal films or van der Waals material substrates[50,51,52] as well as on plasmonic[19,20,21,53] and phononic resonators[31,54,55,56,57]. Although the required substrate structuring might limit imaging applications, the combination of nano-FTIR with polariton launchers and resonators could become a valuable platform for studying strong light-matter interactions at the nanoscale, particularly of infrared light and molecular vibrations[20,31].

In summary, we first demonstrated the crucial role of highly reflecting substrates for enhancing the infrared nanospectroscopy signals of thin molecular layers. Au substrates were found to provide an about one order of magnitude enhancement of nano-FTIR spectral signatures as compared to $CaF_2$ substrates. Exploiting tip-induced surface phonon polariton excitation in polar crystal substrates (here quartz), a further enhancement by a factor of 3 can be achieved. An even further enhancement could be achieved by implementing additional tip illumination which includes propagating surface polaritons, altogether boosting the nano-FTIR spectroscopic signal of molecular vibrations by nearly one order of magnitude compared to nano-FTIR employing Au substrates. Polariton-enhanced nano-FTIR spectroscopy - eventually in combination with infrared-resonant tips[58] - could open promising avenues for molecular vibrational spectroscopy of very thin layers of analytes, monolayers or maybe even single molecules. Finally, we speculate that polariton-enhanced near-field coupling could also boost the sensitivity in tip-enhanced photothermal infrared nanospectroscopy[59] and photo-induced force microscopy[60,61].

**Acknowledgements**
The authors financial support from the Spanish Ministry of Science, Innovation and Universities (national projects MAT2015-65525-R, RTI2018-094830-B-100, and the project MDM-2016-0618 of the Marie de Maeztu Units of Excellence Program). This work also has received funding from the European Union's Horizon2020 research and innovation programme under the Marie Sklodowska-Curie grant agreement No. 721874 (SPM2.0).

**Competing interests:** R.H. is co-founder of Neaspec GmbH, a company producing scattering-type scanning near-field optical microscope systems, such as the one used in this study. The remaining authors declare no competing interests.

**Methods**:
Nano-FTIR spectra are calculated using the finite dipole model[23], with an extension to multi-layered samples[24]. In the model, the tip is described as a prolate spheroid (with a major half axis of length L and an apex radius R) that interacts with the sample made of multiple layers, each of them being characterized by their dielectric function. The employed model parameters are R=25nm, L=150nm, tapping amplitude A=37nm and g=0.6+i0.2 (in the model, the empirical factor g describes the portion of charge induced in the tip, that is relevant for near-field interaction[23]).

# Supporting Information

# Substrate matters: Surface-polariton enhanced infrared nanospectroscopy of molecular vibrations

Marta Autore[1], Lars Mester[1], Monika Goikoetxea[1], R. Hillenbrand[1,2]

[1] CIC nanoGUNE, 20018 Donostia-San Sebastián, Spain
[2] IKERBASQUE, Basque Foundation for Science, 48013 Bilbao, Spain

r.hillenbrand@nanogune.eu

**S1. Variation of enhancement factor *C* for different normalization procedures**

We performed simulations to corroborate the validity of our procedure that is used for determining the molecular vibrational contrast *C*. We recall that in Fig. 2 to 4 of the main text the pure quartz spectrum was scaled to match the spectrum of PEO on quartz in a spectral region outside the PEO vibrational spectral range (black curve in Fig. S1a). We then subtracted the quartz spectrum from the PEO on quartz spectrum to obtain $\Delta s_4$ and *C*. This procedure is most straight forward, as it does not require demanding fitting of the experimental spectra. However, it neglects the spectral shift of the polaritonic tip-substrate resonance that occurs when the tip-substrate distance is increased[1,2]. We validate the reliability of this procedure by calculating the quartz spectrum when a thin layer is on top of the quartz surface, with a layer thickness corresponding to that of the PEO layer and a permittivity corresponding to that of the PEO layer without the molecular vibrational feature. We assume different values for the PEO permittivity: $\varepsilon_{PEO}$ = 2.0 (green spectrum in Fig. S1a), 1.7 (dashed green spectrum) and 1.4 (dotted green spectrum). The higher and lower considered values correspond to the permittivity at 1050 cm$^{-1}$ and at 1300 cm$^{-1}$, while the third one is their average. That way the spectral shift is considered and can be observed by comparing the black and green spectra in Fig. S1a. Subtracting the differently normalized quartz spectra (black and green) from the PEO on quartz spectrum (red), we obtain the spectral contrast $\Delta s_4$ (not shown) from which we can determine *C*, the latter shown in Fig. S1b. The variation of *C* (green shaded area in Fig. S1b) is significantly smaller than the 300% enhancement of the spectral contrast of PEO on quartz compared to the spectral contrast of PEO on gold. For smaller PEO thicknesses the variation of *C* reduces, while for larger PEO thicknesses it increases. This observation can be better appreciated by plotting the ratio $C_{c-SiO2}(d)/C_{Au}(d)$ in Fig S1c. We note that polariton-resonant tip-substrate coupling will be most beneficial for thin layers below 10 nm thickness, where the variation caused by our normalization procedures reduces.

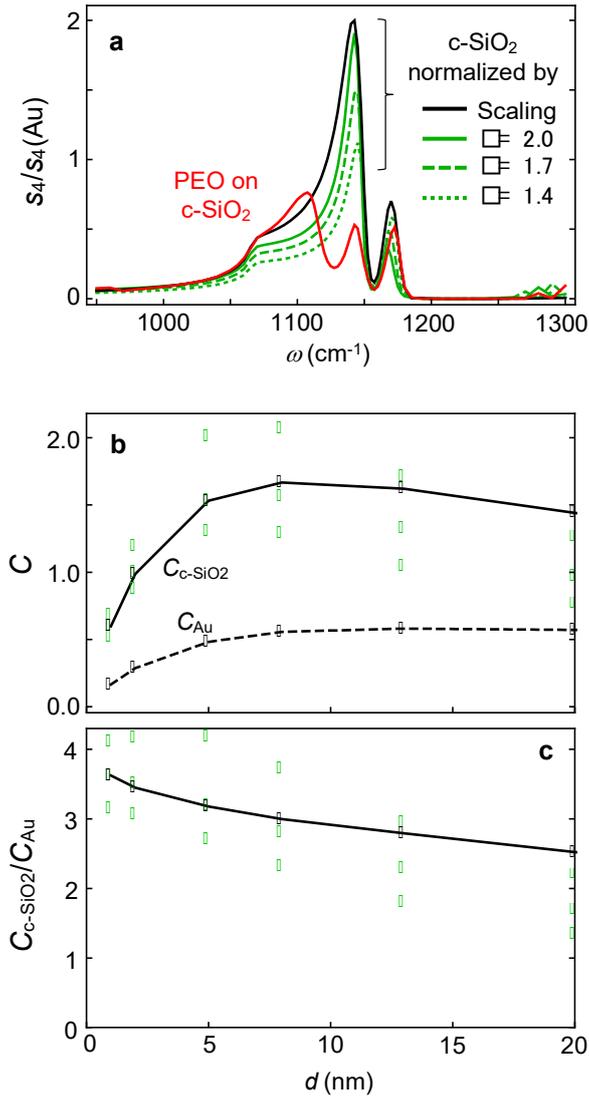

**Figure S1. Variation of enhancement factor *C* for different normalization procedures.** (a) Calculated nano-FTIR amplitude spectra of 13 nm-PEO on c-SiO$_2$ (red), in comparison with c-SiO$_2$ spectra normalized in different ways: Scaling as in the main text (black) and by calculating the c-SiO$_2$ spectrum with a 13 nm-thick dielectric layer of permittivity $\varepsilon_{PEO}$ = 2.0 (solid green), $\varepsilon_{PEO}$ = 1.7 (dashed green) and $\varepsilon_{PEO}$ = 1.4 (dotted green). The latter three curves account for spectral shifts of the phonon resonance due to increased tip-substrate distance. (b) Maximum spectral contrast for differently thick PEO layers on quartz and Au, $C_{c\text{-}SiO2}$ and $C_{Au}$, obtained for the different normalization procedures of the pure c-SiO$_2$ spectra according to Fig S1a. (c) Ratios $C_{c\text{-}SiO2}/C_{Au}$.

## S2. Nano-FTIR phase spectra of PEO on quartz substrate

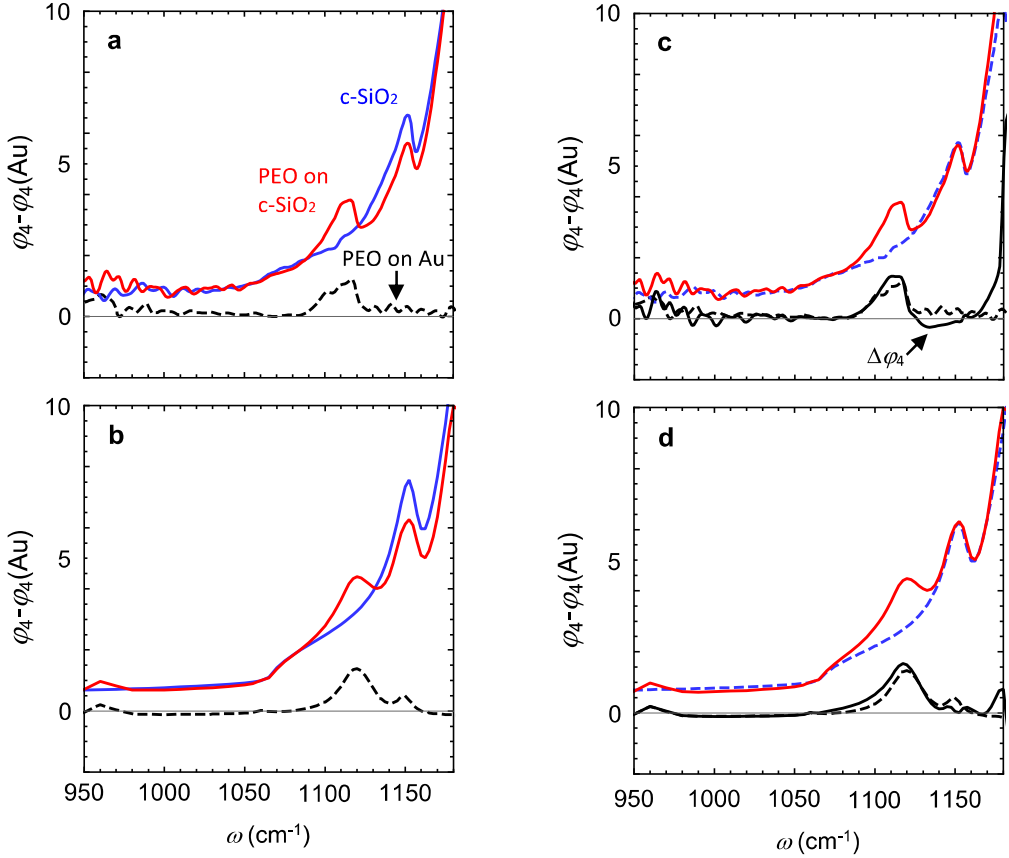

**Figure S2. nano-FTIR phase spectra of PEO on quartz substrate.** (a) Experimental and (b) calculated nano-FTIR phase spectra (4th harmonics, normalized to gold spectrum) of quartz (blue), 13 nm-thick PEO on quartz (red) and 13 nm-thick PEO on Au (black dashed curve). Panels (c) and (d) show the quartz spectra normalized by scaling (blue dashed) and the spectral contrast $\Delta\varphi_4 = \varphi_4(\text{PEO on c-SiO}_2) - \varphi_4(\text{c-SiO}_2)$ (black solid curve). The maximum spectral phase contrast is barely enhanced by the c-SiO$_2$ substrate compared to the Au substrate.

## S3. Near-field spectra of PEO on quartz as function of distance from the Au edge

To demonstrate the influence of the edge-launched surface phonon polariton on quartz on the spectral contrast $C$, we measured nano-FTIR spectra at various distances from the gold edge. The spectra recorded on pure quartz (obtained when the tip is positioned in between PEO islands, Fig. S3c) clearly show a modulation of the peak maxima with increasing distance from the Au edge. The peak height agrees well with the behaviour of the s-SNOM amplitude signal shown in Fig. S3b. We then recorded the spectra of PEO on quartz, at the same distances from the Au edge obtained when the tip is positioned on top of the PEO islands, Fig. S3d) and calculated the spectral contrast $\Delta s_4$ (Fig. S3e). We find that the spectral contrast indeed follows the local field above the quartz surface, which is determined by the interference of the incident field and the field of the surface phonon polariton launched at the gold edge.

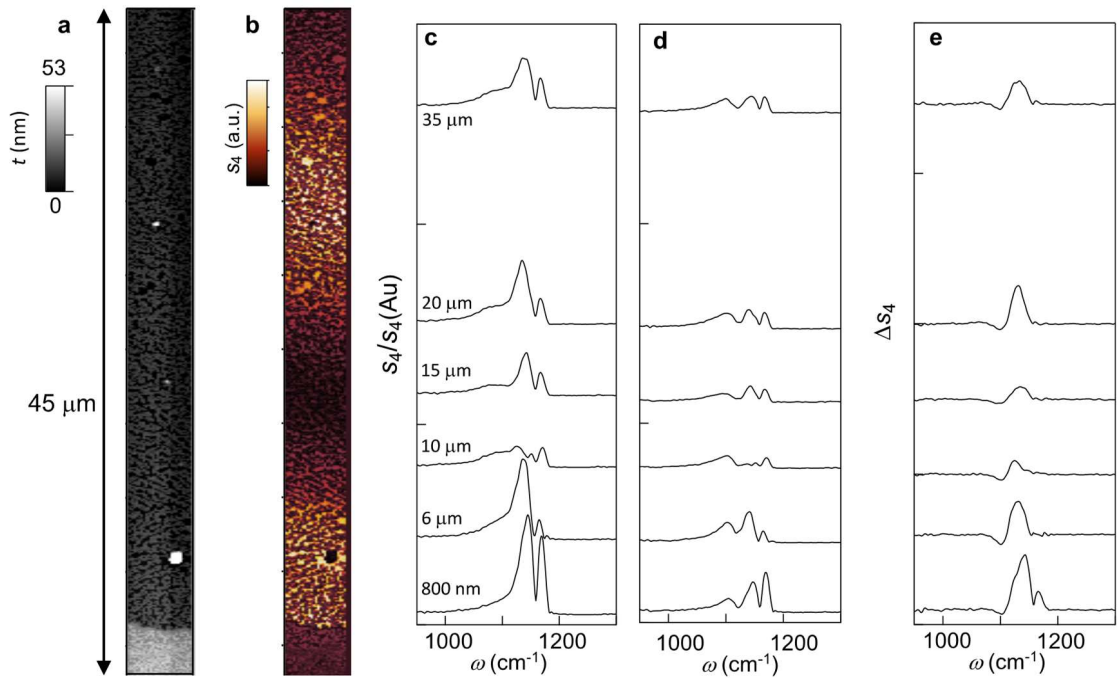

**Figure S3. s-SNOM images and nano-FTIR phase spectra of PEO on quartz substrate close to a gold edge.** (a) AFM topography image and (b) infrared amplitude image (recorded at 1114 cm$^{-1}$) of a large area of PEO on quartz close to a gold edge (seen at the bottom). The images are the same as the ones shown in Fig. 3 of the main text. (c and d) nano-FTIR amplitude spectra of quartz (c) and PEO on quartz (d), recorded at several distances (provided by the numbers labeling the spectra) from the gold edge. (e) spectral contrast $\Delta s_4$ calculated at each distance, following the procedure described in the main text.